\documentclass[fleqn,usenatbib]{mnras}

\usepackage{newtxtext,newtxmath}

\usepackage[T1]{fontenc}
\usepackage{ae,aecompl}

\usepackage{graphicx}	
\usepackage{amsmath}	
\usepackage{amssymb}	

\usepackage{color}
\usepackage{bm}
\usepackage{xspace}
\usepackage{pgffor}
\usepackage{upgreek}
\usepackage{nicefrac}
\usepackage[normalem]{ulem}
\usepackage[single=true]{acro}

\newcommand{\rikkyo}{{Department of Physics, Rikkyo University, Nishi-Ikebukuro 3-34-1, Toshima-ku, Tokyo 171-8501, Japan}}

\newcommand{\eg}{{\itshape e.g.}\xspace}

\newcommand\ie{{\itshape i.e.}\xspace}
\newcommand{\erg}{\ensuremath{\mathrm{erg}}}
\newcommand{\ergs}{\ensuremath{\mathrm{erg\,s^{-1}}}}
\newcommand{\ergscm}{\ensuremath{\mathrm{erg\,s^{-1}\,cm^{-2}}}}

\newcommand{\kms}{\ensuremath{\mathrm{km\,s^{-1}}}}
\newcommand{\cccs}{\ensuremath{\mathrm{cm^{3}\,s^{-1}}}}

\usepackage{pgffor}
\makeatletter
\def\dif{\@ifnextchar[{\@with}{\@without}}
\def\@with[#1]#2{
  \ensuremath{\frac{\foreach \x in {#2}{\mathrm{d}\hskip.08em\x\,}}{\foreach \x in {#1}{\mathrm{d}\hskip.08em\x\,}}}
}
\def\@without#1{
  \ensuremath{%
    \ifx\hfuzz#1\hfuzz
    \mathrm{d}
    \else
    \foreach \x in {#1}{\mathrm{d}\hskip.08em\x\,}
    \fi
    }
}
\def\mitya{\@ifnextchar[{\@mwith}{\@mwithout}}
\def\@mwith[#1]#2{
  {\color{blue} \bfseries #2}\footnote{Comment: #1}
}
\def\@mwithout#1{
  {\color{blue} \bfseries #1}
}
\makeatother

\usepackage{xifthen}
\newcommand{\ism}[1][]{\ensuremath{{}_\textsc{ism\ifthenelse{\equal{#1}{}}{}{,\tiny\,#1}}}}
\newcommand{\ns}[1][]{\ensuremath{{}_\textsc{ns\ifthenelse{\equal{#1}{}}{}{,\tiny\,#1}}}}
\newcommand{\win}[1][]{\ensuremath{{}_\textsc{w\ifthenelse{\equal{#1}{}}{}{,\tiny\,#1}}}}
\newcommand{\hi}{\ensuremath{{}_{\scriptscriptstyle  HI}}}
\newcommand{\hy}{\ensuremath{{}_{\scriptscriptstyle  H}}}
\newcommand{\ha}{\ensuremath{{}_{\scriptscriptstyle H\alpha}}}

\DeclareAcronym{snr}{
  short = SNR ,
  long  = supernova remnant ,
  class = astro ,
  first-style = default
}
\DeclareAcronym{pwn}{
  short = PWN ,
  long  = pulsar wind nebula ,
  short-plural = e ,
  long-plural = e ,
  class = astro ,
  first-style = default
}
\DeclareAcronym{cr}{
  short = CR ,
  long  = cosmic ray ,
  class = astro ,
  first-style = default
}
\DeclareAcronym{ns}{
  short = NS ,
  long  = neutron star ,
  class = astro ,
  first-style = default
}
\DeclareAcronym{ism}{
  short = ISM ,
  long  = interstellar medium ,
  class = astro ,
  first-style = default
}
\DeclareAcronym{hd}{
  short = HD ,
  long  = hydrodynamic ,
  class = hydro ,
  first-style = default 
}

\DeclareAcronym{mhd}{
  short = MHD ,
  long  = \acifused{hd}{magneto-HD}{magnetohydrodynamic} ,
  class = hydro ,
  first-style = default 
}
\DeclareAcronym{rmhd}{
  short = RMHD ,
  long  = \acifused{mhd}{relativistic MHD}{\acifused{hd}{relativistic magneto HD}{relativistic magnetohydrodynamic}} ,
  class = hydro ,
  first-style = default
}
\DeclareAcronym{1d}{
  short = 1D ,
  long  = one-dimensional ,
  class = hydro ,
  first-style = short
}
\DeclareAcronym{2d}{
  short = 2D ,
  long  = two-dimensional ,
  class = hydro ,
  first-style = short
}
\DeclareAcronym{3d}{
  short = 3D ,
  long  = three-dimensional ,
  class = hydro ,
  first-style = short
}
\DeclareAcronym{cd}{
  short = CD ,
  long  = contact discontinuity ,
  class = hydro ,
  first-style = default
}
\DeclareAcronym{ts}{
  short = TS ,
  long  = termination shock ,
  class = hydro ,
  first-style = default
}
\DeclareAcronym{fs}{
  short = FS ,
  long  = forward shock ,
  class = hydro ,
  first-style = default
}
\DeclareAcronym{rs}{
  short = RS ,
  long  = reverse shock ,
  class = hydro ,
  first-style = default
}
\DeclareAcronym{eos}{
  short = EOS ,
  long  =  equation of state ,
  long-plural-form = equations of state ,
  class = hydro ,
  first-style = default
}
\DeclareAcronym{kh}{
  short = KH ,
  long  =  Kelvin~--~Helmholtz ,
  class = hydro ,
  first-style = default
}

\DeclareAcronym{pic}{
  short = PIC ,
  long = particle-in-cell ,
  first-style = default
}

\title[Fast moving pulsars as probes of ISM]{Fast moving pulsars as probes of interstellar medium}
\author[M.V. Barkov et al.]{
Maxim V. Barkov,$^{1,2,3,4}$\thanks{E-mail: mbarkov@purdue.edu}
Maxim Lyutikov$^{1}$ and Dmitry Khangulyan$^{5}$
\\
$^{1}$ Department of Physics and Astronomy, Purdue University, West Lafayette, IN 47907-2036, USA\\
$^{2}$ Astrophysical Big Bang Laboratory, RIKEN, 351-0198 Saitama, Japan \\
$^{3}$ Max-Planck-Institut f\"ur Kernphysik, Saupfercheckweg 1, 69117 Heidelberg, Germany \\
$^4$ Institute of Astronomy, Russian Academy of Sciences, Moscow, 119017 Russia\\
$^{5}$\rikkyo
}

\date{Accepted XXX. Received YYY; in original form ZZZ}

\pubyear{2015}

\begin{document}
\label{firstpage}
\pagerange{\pageref{firstpage}--\pageref{lastpage}}
\maketitle

\begin{abstract}
  Pulsars moving through \ac{ism} produce bow shocks detected in hydrogen H$\alpha$ line { emission}. The morphology of the bow shock nebulae allows one to probe the properties of \ac{ism} on scales $\sim 0.01$ pc and smaller. We performed 2D \ac{rmhd} modeling of the pulsar bow shock and simulated the corresponding H$\alpha$ {emission} morphology. We find that even
  a mild spatial inhomogeneity of \ac{ism} density, $\delta\rho/\rho \sim 1$, leads to significant variations of the shape of the shock seen in H$\alpha$ line { emission}. We successfully reproduce the morphology of the Guitar Nebula. We infer   quasi-periodic density variations in the warm component of \ac{ism} with characteristic length { of}  $\sim0.1$~pc. Structures of this scale might be also responsible for {the} formation of the fine features seen at the forward shock { of} Tycho \ac{snr} in X-rays.  Formation of such short periodic density structures in the warm component of \ac{ism} is puzzling, { and bow-shock nebulae provide unique probes to study this phenomenon.}
\end{abstract}

\begin{keywords}
radiation mechanisms: thermal -- hydrodynamics -- stars: neutron  -- pulsars: individual: B2224+65
\end{keywords}



\section{Introduction}
\label{intro}
\acresetall

Pulsars  produce ultra-relativistic winds that create  Pulsar Wind Nebulae \citep[PWNe,\acuse{pwn}][]{reesgunn,2006ARA&A..44...17G,2008AIPC..983..171K,2015SSRv..191..391K,2017SSRv..207..175R}. Many fast moving  pulsars quickly escape the host  supernova remnant  \citep[for a recent review see][]{kpkr17}.  The proper speed of such pulsars vary in {\bf a} wide range from a few kilometers per second to more than a thousand kilometers per second. Typically, these velocities are much larger   than the  sound speeds in the \ac{ism}, $c_{s,}\ism \sim 10-100\,\kms$. 

{The interaction of a fast moving pulsar wind with the \ac{ism} produces a bow-shock nebula with extended tail. The X-ray emission of the bow-shock was discussed by \citet{2019MNRAS.484.4760B,2019MNRAS.488.5690O}; the filamentary structure created by non-thermal particles escaping from PWN was researched by \citet{ban08,2019MNRAS.489L..28B,2019MNRAS.485.2041B}. }

In the apex part of the \ac{pwn} the \ac{ism} ram pressure confines the pulsar wind  producing two shocks -- forward shock in the \ac{ism} and the reverse/\ac{ts} in the pulsar wind; 
the two shock are separated by the contact discontinuity (CD).
This picture is similar to the structure formed by interaction of the  Solar wind with the Local \ac{ism} 
\citep[see][for review]{zank99,2017ApJ...845....9P}.

Early works on the structure of the bow shock nebulae \citep[\eg][]{2005A&A...434..189B} predicted { a} smooth morphology. In contrast,  the observations show large morphological variations \citep{2017JPlPh..83e6301K}, both of the \ac{pwn} part (regions encompassing the shocked  pulsar wind)  and the shape of the bow shock (in the \ac{ism} part of the  interaction region).  Especially puzzling is the Guitar Nebula, which shows what can be called ``closed in the back'' morphology: shocks, delineated by the $H\alpha$ emission, bend/curve towards the tail, not the head part.  

Variations of the \ac{pwn} part of the interacting flow are likely  due to the internal dynamics of the shocked pulsar wind: mass loading \citep{2015MNRAS.454.3886M,2018MNRAS.481.3394O}, as well as fluid and current-driven instabilities \citep{2019MNRAS.484.4760B} can induce larger variations in the shape of the confined pulsar wind. (We note here that so far no simulations were able to catch the long term dynamics in the tails, on scales much longer than the stand-off distances.)
 
One of the goals of the present study is to investigate whether the variations of the \ac{pwn} part of the flow  can affect the shape of the forward shock \citep[\eg, as discussed by][]{Kerkwijk08}. Our simulations indicate that {the} internal dynamics of the \ac{pwn} part does not affect the bow shock in { any} appreciable way. The basic reason is that {\bf the} mildly relativistic flow within the \ac{pwn} tail quickly advects { downstream} all the perturbations, see \S \ref{sec:res}.
 
We  then resort to external perturbations to produce the morphological variations of the $H\alpha$ { emission} maps. We target the Guitar Nebula  as the extreme example of { a}  nontrivial morphology of the forward shock. The most surprising morphological feature of the Guitar {Nebula} is a sequence of ``closed in the back'' shocks \citep{2007MNRAS.374..793V,2017MNRAS.464.3297Y,2019MNRAS.484.1475T}. 
Several fast moving pulsars form nebula{\bf e} with  { $H\alpha$ line} emission \citep{2014ApJ...784..154B}, which show  sequences of ``closed in the back'' shocks similar to those seen in the Guitar Nebula.
As we demonstrate, the ``closed in the back'' morphology is a { result of the} projection effect (due to particular combination of the line of sight) and of the physical properties {of the system} (pulsar's velocity and \ac{ism} density variations).

\section{Physics of hydrogen ionisation}
\label{sec:physics}

{ A} fast moving pulsar form { a } bow shock { that} heats the electrons and ions {from \ac{ism}}. Hot electrons ionize neutral atoms and thermalize with them. 
The pressure balance between the pulsar wind and the \ac{ism} gives the stand-off distance $r_s$
\begin{equation}
  r_s =\sqrt{\frac{L\win }{ 4 \uppi c  \rho\ism \varv\ns^2}} = 4 \times 10^{16}  L\win[36]^{1/2} \, n\ism[0]^{-1/2} \varv\ns[7.5]^{-1} \, {\rm cm}\,,
\label{eq:rs}  
\end{equation}
where $L\win$ is {pulsar spin-down luminosity}, $\varv\ns$ is the velocity of pulsar,  $\rho\ism=\mu m_p n\ism$ is \ac{ism} mass and number  density (here $m_p$ is mass of proton, { $\mu = 1.4$ is chemical composition factor}).  We use the following normalization agreement: $A=10^x A_x\rm\,cgs\,units$. {The pass-though time, which is the characteristic time for crossing the stand-off distance, is} 
\begin{equation}
  t_s = r_s/\varv\ns = 1.3 \times 10^{9}  L\win[36]^{1/2} \, n\ism[0]^{-1/2} \varv\ns[7.5]^{-2} \, {\rm s}\,.
\label{eq:ts}  
\end{equation}

Let us  next estimate  the ionization and recombination rates for hydrogen  at the  shock front. Following the work \citep{1997A&A...321..672R} the ionization rate is 
\begin{equation}
    \dif[t]{x\hi} = n_e \left[c_r (1-x\hi) -c_i x\hi \right]   
\label{eq:dfndt}
\end{equation}
here  $n_e=n\hy\left[(1-x\hi)+0.001\right]$ is electron concentration\footnote{{The value 0.001 takes into account electrons from ionized metals in ISM, see more technical details of ionization process calculations in \citep{1997A&A...321..672R}.}}, $c_i$ and  $c_r$ are ionization and recombination rate coefficients
\begin{equation}
    c_i = 5.8  \times 10^{-11} \sqrt{T_0} \exp{\left(-\frac{1.57\times10^5}{T_0}\right)} \quad\cccs 
\label{eq:ci}
\end{equation}
and
\begin{equation}
    c_r = 2.6  \times 10^{-11} \frac{1}{\sqrt{T_0}} \quad\cccs. 
  \label{eq:cr}
\end{equation}

In the case of a fast moving pulsar the temperature behind the bow shock can be as high as  \(T\simeq 2 \times10^6 \varv\ns[7.5]^2\rm\,K\). At such high temperature{\bf s} ionization processes dominates over recombination ones by far ($c_i \approx 10^{-7}\varv\ns[7.5]$ and $c_r \approx 2\times10^{-14}\varv\ns[7.5]^{-1}$~cm$^{-3}$~s$^{-1}$). 
The ionization time scale can be estimated as 
 \begin{equation}
    t_i =\frac{ 1}{c_i n_e} \sim  10^{7} n_e^{-1} \varv\ns[7.5]^{-1} \mbox{ s}. 
  \label{eq:ti}
\end{equation} 
Comparing Eq.~(\ref{eq:ts}) and Eq.~(\ref{eq:ti}) we can see $t_i\ll t_s$, it means that the ionization is very fast and thickness of the {ionization shell} should be small. {In contrast, the recombination time scale \(t_r\sim5\times10^{13}\varv\ns[7.5] \) is much { longer as compared} to the pass-through time scale.  Some atoms ionization process is accompanied by excitation and emission of \(H\alpha\) line. The most efficient condition of \(H\alpha\) line emission is near $x\hi\approx 0.5$.  As the ionization proceeds in a significantly more compact region compare to recombination one, it should be responsible for generation of the most distinct line emission. } Thus, H$\alpha$ emission is produced in a thin shell right behind the bow shock. Only for slow pulsars, with velocity below $10^7$~cm~s$^{-1}$, the ionization shell thickness can be comparable with the bow shock size.

Using the algorithm {suggested by} \citet{1997A&A...321..672R}, hydro-dynamical data as density, pressure and $x\hi$ (see section~\ref{sc:numset}) allow us to calculate the {excitation} rate of hydrogen and intensity of H$\alpha$ line { emission} as
\begin{equation}
    \dot{e}\ha = n\hy n_e x\hi j\ha \qquad \mbox{\erg\,\cccs} 
    \label{eq:eHa}
\end{equation}
 and emissivity coefficient 
\begin{equation}
    j\ha = \frac{\hbar^2\sqrt{2\uppi}}{k_B^{1/2}m_e^{3/2}}q_{12}
    \frac{h\nu\ha}{1+\left(q_{21}/A_{21}\right)} \quad \mbox{\erg\, cm}^{-3} \mbox{ s}^{-1},
    \label{eq:jHa}
\end{equation}
here
\begin{equation}
    q_{21}=\frac{8.6\times10^{-6}}{T_0^{1/2}}\frac{\Omega_{21}}{g_2}
    \label{eq:q21}
\end{equation}
and
\begin{equation}
    q_{12}=\frac{8.6\times10^{-6}}{T_0^{1/2}}\frac{\Omega_{12}}{g_1}\exp{\left(-\frac{h\nu\ha}{k_B T}\right)},
    \label{eq:q12}
\end{equation}
where parameters $g_1$, $g_2$ and $\Omega_{21}=\Omega_{12}$ are tabulated in PLUTO code\footnote{Link http://plutocode.ph.unito.it/index.html}.


\section{Method}
\label{sc:method}

We implement a two-step procedure to calculate the expected morphology of the H$\alpha$ emission.
First, as described  in \S \ref{sc:numset} we perform relativistic \ac{mhd} simulations of a pulsar moving through {inhomogeneous} \ac{ism}. These calculation{\bf s} {define} the shape of the bow shock { and values of all relevant hydrodynamic parameters}. Next, \S \ref{sc:hamaps}, we performed post-processing of the {hydrodynamic} data to calculate H$\alpha$ emissivity maps.

\subsection{Numerical  Setup}
\label{sc:numset}
We performed a number of \ac{2d} \ac{rmhd} simulations of the interaction of relativistic pulsar wind with external medium.  The simulations were performed using a \ac{2d}  geometry in Cylindrical coordinates using the {\it PLUTO} code \citep{mbm07}.  Spatial parabolic interpolation, a 2${}^{\rm nd}$ order Runge-Kutta approximation in time, and an HLLC Riemann solver were used \cite{2005MNRAS.364..126M}.  {\it  PLUTO} is a modular Godunov-type code entirely written in C and intended mainly for astrophysical applications and high Mach number flows in multiple spatial dimensions.  The simulations were performed on CFCA XC50 cluster of national astronomical observatory of Japan (NAOJ).  The flow has been approximated as an ideal, relativistic adiabatic gas, one particle species, and polytropic index of 4/3\footnote{{We choose this value to describe relativistic flow accurately. Our previous simulations have shown small sensitivity of flow general structure to a change of the polytropic index from 4/3 to 5/3 or adaptation of Taub EOS \citep{2019MNRAS.484.4760B, 2019MNRAS.485.2041B}}}. The size of the domain is $R \in [0, 20 a]$, $Z \in [-3 a, 100 a] $ where the length unit $a=10^{16}$~cm (the initial \ac{ism} velocity is directed along $Z$-axis). To have a good resolution in the central region and long { the} tail zone we use { a non-uniform} resolution in the computational domain. {For ``high-resolution'' simulations we adopted the following grid configuration. Number of the radial cells in a high resolution uniform grid in the central region (for $R \in [0, 1 a] $) is $N_{\rm R, in} = 520$. The outer { non-uniform} grid (for region with $R \in [1 a, 20 a]$) has $N_{\rm R, out} = 1560$ radial cells. In the Z direction we also adopted a non-uniform grid with three different zones. The inlet part ($Z \in [-3 a, -1 a]$) has  $N_{\rm Z, out-} = 520$ { non-uniform} grid. The  central part (for $Z \in [-1 a, 1 a] $) has a uniform grid with $N_{\rm Z, in} = 1040$ cells, and the { non-uniform} tail grid (for $Z \in [1 a, 100 a]$) contains $N_{\rm Z, out+} = 4680$ cells.} {The} grid parameters in {``low-resolution''} models are { the} following:  { $N_{\rm R, in} = 130$ and $N_{\rm R, out} = 390$ for the radial direction; $N_{\rm Z, out-} = 130$, $N_{\rm Z, in} = 260$, and $N_{\rm Z, out+} = 1170$ for the Z-direction.}  See Table~\ref{tab:models} for other parameters.

We used { a} so called simplified non-equilibrium cooling (SNEq) block of {\it PLUTO} code \citep[see details in ][]{1997A&A...321..672R}. Which allows to calculate the fraction of neutral atoms of hydrogen $x\hi$ in \ac{ism} taking into account ionization and recombination processes. Moreover, the radiation loses for 16 lines (like Ly~$\alpha$, HeI (584+623),  OII (3727) etc.) are  included in the energy equation. 

\subsection{H$\alpha$ emissivity maps}
\label{sc:hamaps}

To obtain H$\alpha$ brightness maps, Eq.~\eqref{eq:eHa} needs to be integrated over the line of sight. These calculation were performed at the postprocessing stage on a workstation-class PC. Despite the simplicity of the numerical procedure, obtaining of high-resolution synthetic maps required quite a large amount of computations. The character of required calculations, \ie integration over the line of sight, implies a case for effective GPU computing. We utilized  {\it PyCUDA}, a {\it CUDA} API implementation for {\it python}, which allowed us to profit python visualization library {\it matplotlib} and get access to high-performance GPU computations, which were performed on {\it NVIDIA TITAN V} card.  The synthetic maps were obtained in Cartesian coordinate system, with Z-axis directed towards the observer. We adopted a uniform XY-grid with resolution of \(1536\times1024\), which is sufficient for {the} comparison with observations. The symmetry axis of the \ac{mhd} cylindrical box is assumed to locate in the XZ-plane and to make angle \(\theta\) with X-axis (``viewing angle''). The size of the \ac{mhd} computational box determined the interval of integration over Z-axis. The transformation of the Cartesian coordinates to the cylindrical of the \ac{mhd} box were performed with a recursive procedure that allowed  { some} improvement of the algorithm performance. At the integration points the intensity of H$\alpha$ emission was computed by bilinear interpolation between the most nearby nodes of the \(RZ\) array obtained with {\it PLUTO} simulations. As the dependence of the emission intensity along the line-of-sight is sectionally linear we used trapezoidal integration method. Some further details of the postprocessing script are given in Appendix~\ref{sec:Halpha}.


We simulate interaction of \ac{ns} supersonically moving in the \ac{ism} with {varying} density profile given by
\begin{equation}
    n\ism(z,t) = \frac{n_0}{1-a_\rho \cos\left[\frac{(z-\varv\ns t)}{d\win}\right]},
    \label{eq:nISM}
\end{equation}
here $n_0=1/$cm$^3$ is \ac{ism} concentration, $d\win=3\times10^{17}$~cm wave length, $a_\rho$ is wave amplitude and $\varv\ns$ \ac{ns} speed in \ac{ism}, the values $\varv\ns$ and $a_\rho$ for the models can be found in the Tab.~\ref{tab:models}. In all cases we assume Mach number in the \ac{ism} $M\ism = \varv\ns/c_{s,\,}\ism = 30$.

We initiate the pulsar wind as spherically symmetric with Lorentz factor 4.9  which corresponds to initial Mach number 45. { Anisotropy of the pulsar wind \citep{2002AstL...28..373B} may have certain impact on the morphology of the \ac{pwn}, but the forward shock is much less sensitive to the pulsar wind anisotropy \citep{2019MNRAS.484.4760B}.} We adopt the spindown luminosity for fast and slow models which forms forward termination shock at the distance $r_s=10^{16}$~cm (see eq.\ref{eq:rs})  if $n\ism = 1/$cm$^3$. This value corresponds to $L\win = 5.6\times10^{38}\,\ergs$ for fast and $L\win = 1.4\times10^{36}\,\ergs$ for slow models respectively.

\section{Results}
\label{sec:res}

\subsection{Overall PWN properties}

\begin{table}
	\centering
	\caption{Models parameters. Here presented name of the models, resolution in radial ``R'' and axial ``Z'' directions, neutron star  speed $\varv\ns$, density amplitude variation parameter $a_{\rho}$ and pulsar wind power $L\win$. }
	\label{tab:models}
	\begin{tabular}{lccccc} 
		\hline
	    Name	      & R & Z & $\varv\ns/c$ & $a_{\rho}$ & $L\win[36]$ \\
		\hline
		hr-fast-const & 2080 & 6240 & 0.1 & 0 & 560 \\
		hr-fast-var3 & 2080 & 6240 & 0.1 & 0.5 & 560 \\
		lr-fast-var3 & 520 & 1560 & 0.1 & 0.5 & 560 \\
		lr-{SNEq}-slow-var3 & 520 & 1560 & 0.005 & 0.5 & 1.4 \\
		lr-{SNEq}-slow-var2 & 520 & 1560 & 0.005 & 0.2 & 1.4  \\
		\hline
	\end{tabular}
\end{table}

\begin{figure*}
	\includegraphics[width=\textwidth]{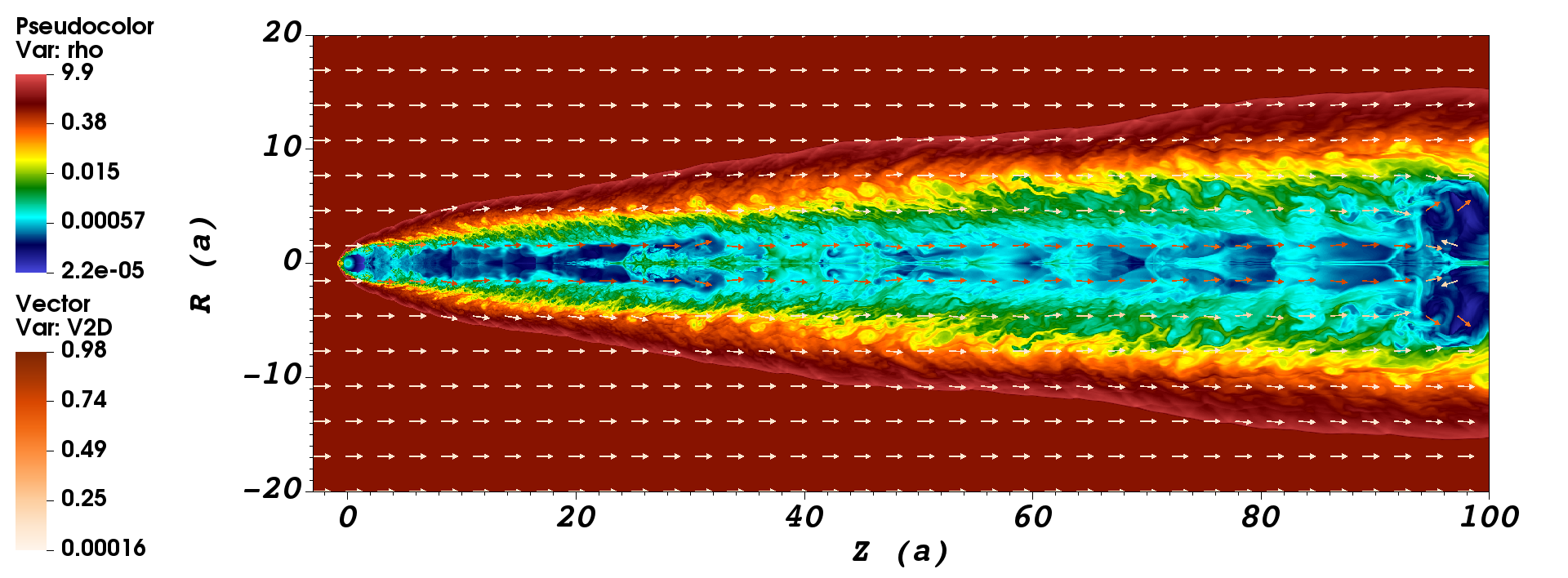}
	\includegraphics[width=\textwidth]{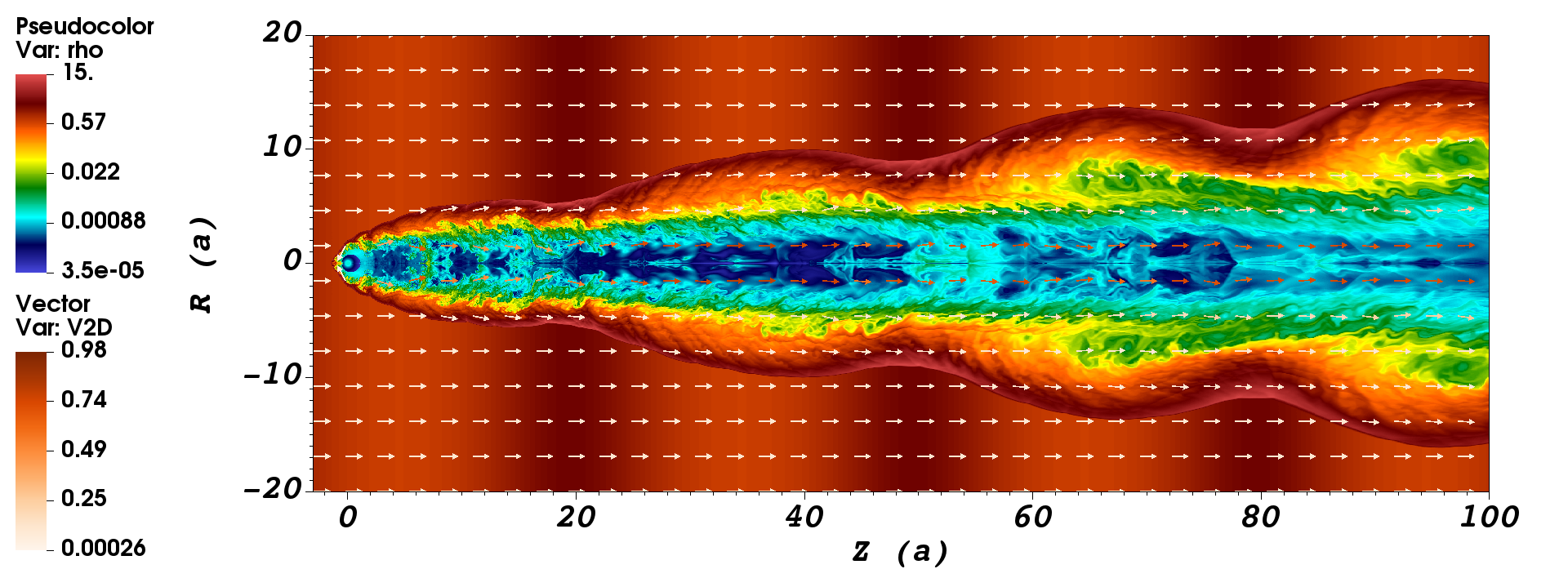}
    \caption{ Color maps of density (log-scale) for hr-fast-const and hr-fast-var3 models  (see Tab.~\ref{tab:models} for detail) in the top and bottom panels, respectively. Arrows show the velocity field, the arrow color defines the flow speed (linear scale, speed of light units).}
    \label{fig:rho_hr}
\end{figure*}

\begin{figure*}
	\includegraphics[width=\textwidth]{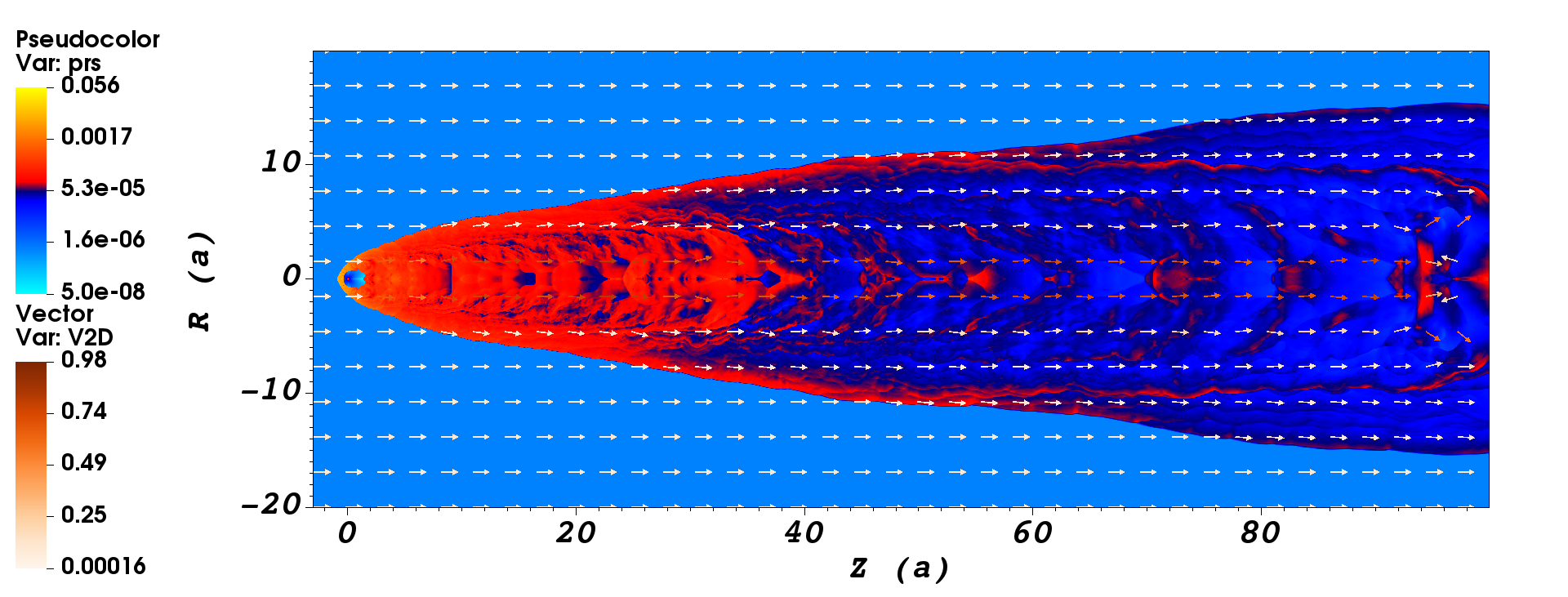}
	\includegraphics[width=\textwidth]{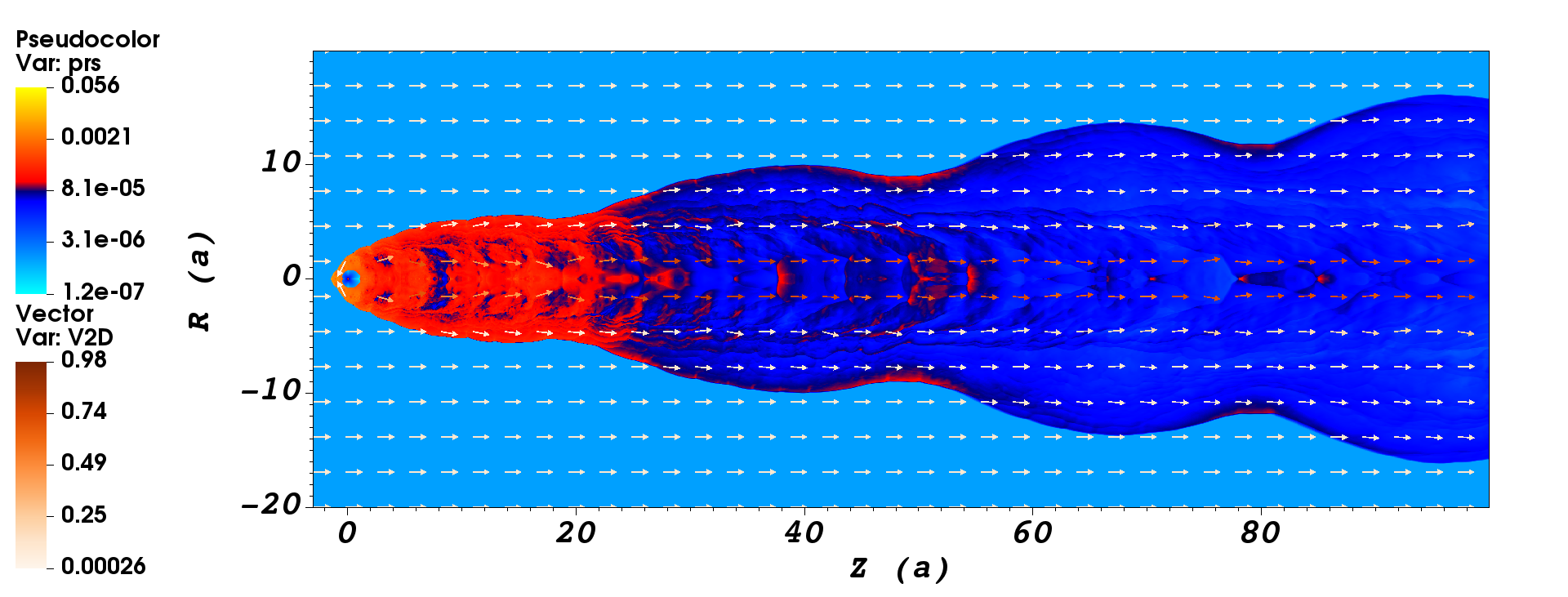}
    \caption{ Color maps of pressure (log-scale) for hr-fast-const and hr-fast-var3 models  (see Tab.~\ref{tab:models} for detail) in the top and bottom panels, respectively. Arrows show the velocity field, the arrow color defines the flow speed (linear scale, speed of light units).}
    \label{fig:prs_hr}
\end{figure*}

\begin{figure*}
	\includegraphics[width=\textwidth]{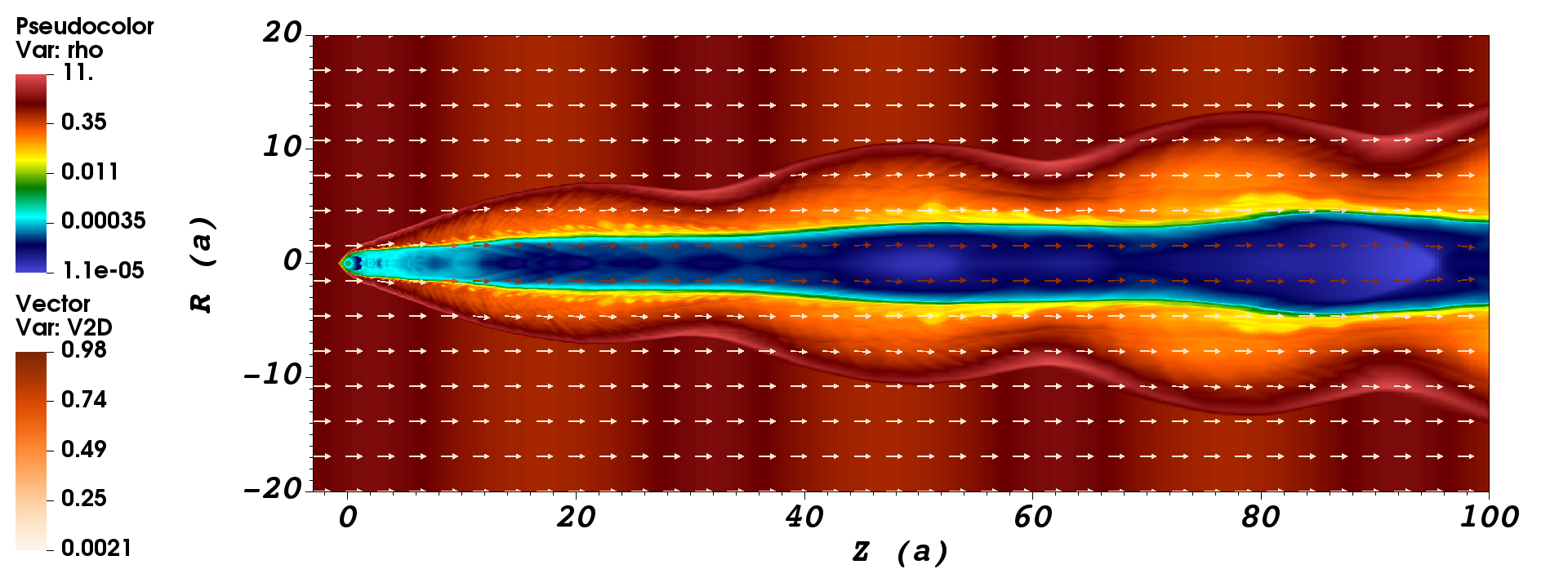}
	\includegraphics[width=\textwidth]{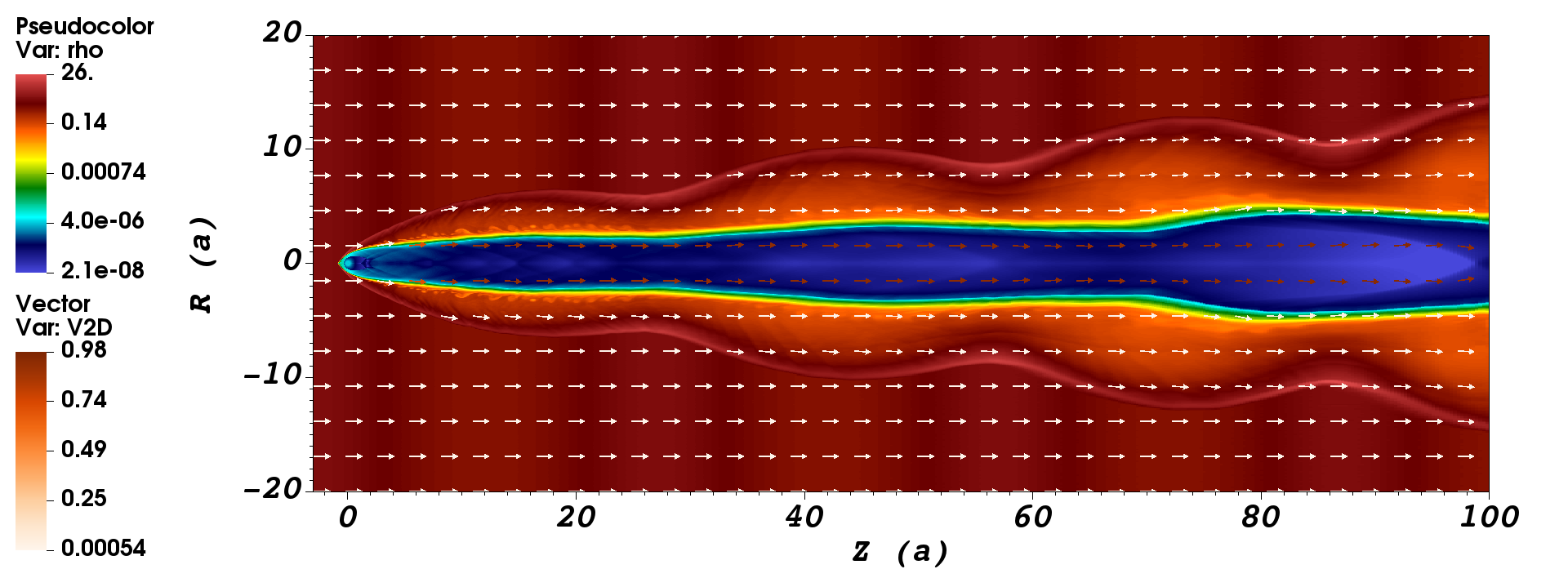}
	\includegraphics[width=\textwidth]{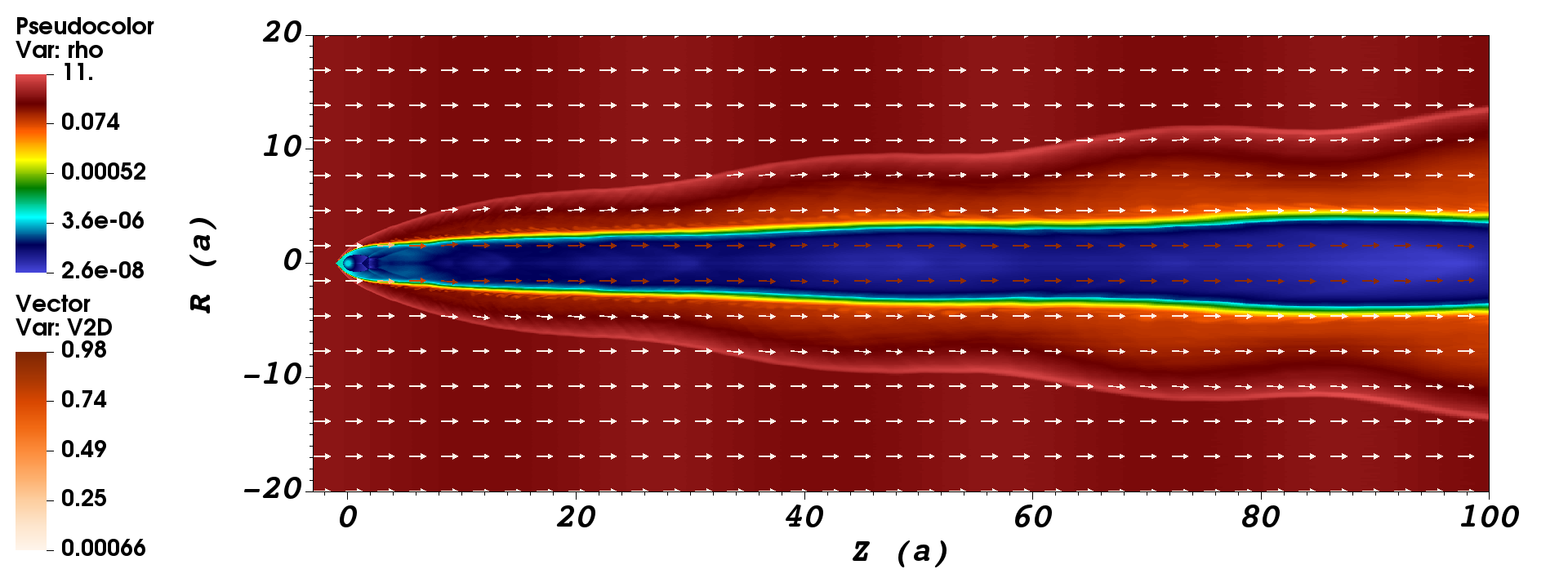}
	\caption{ Color maps of density (log-scale) for three ``low-resolution'' models: lr-fast-var3 (top panel), lr-SNEq-slow-var3 (middle panel),  and lr-SNEq-fast-var2 (which has a smaller variation of \ac{ism} density, bottom panel). Arrows show the velocity field, the arrow color defines the flow speed (linear scale, speed of light units).}
    \label{fig:rho_lr}
\end{figure*}

\begin{figure*}
	\includegraphics[width=\textwidth]{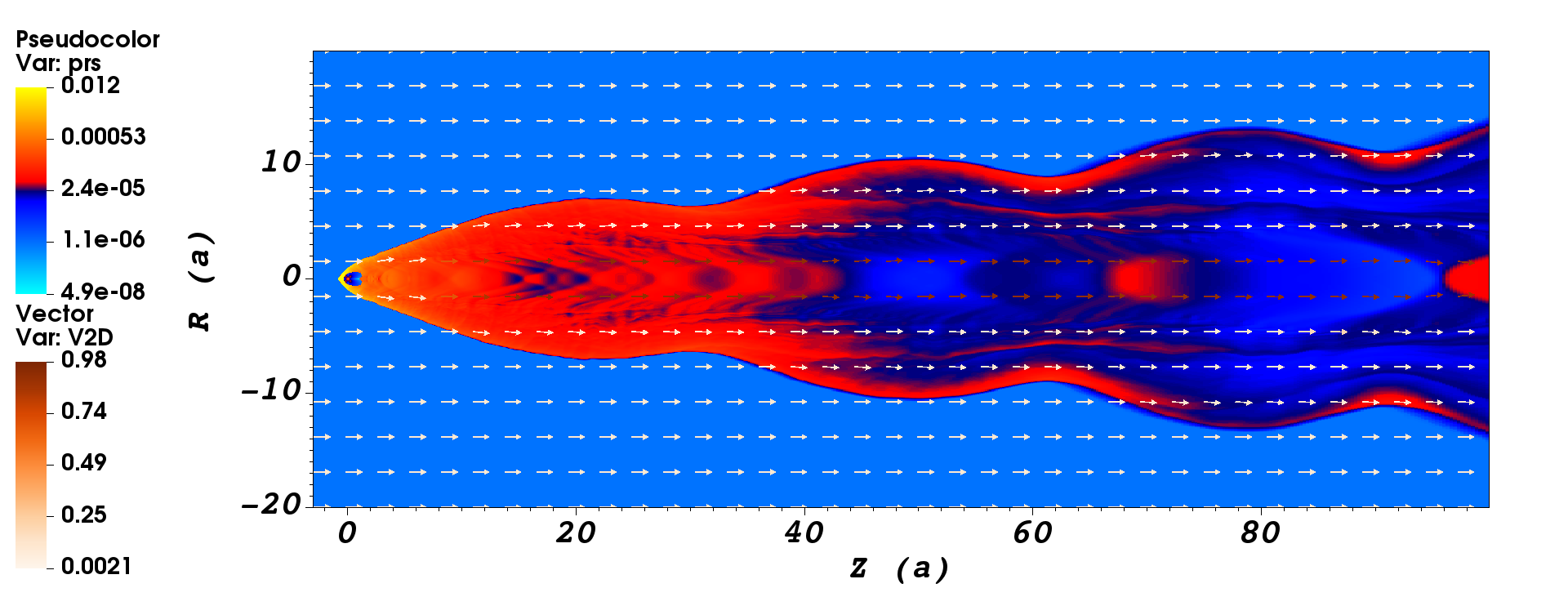}
	\includegraphics[width=\textwidth]{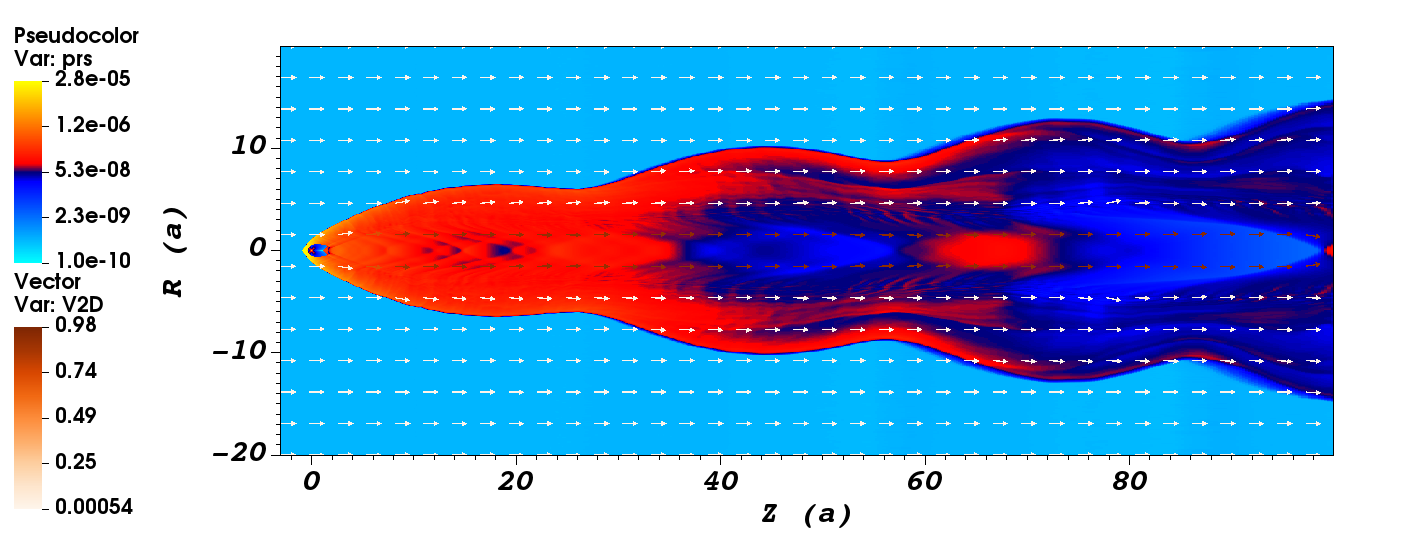}
	\includegraphics[width=\textwidth]{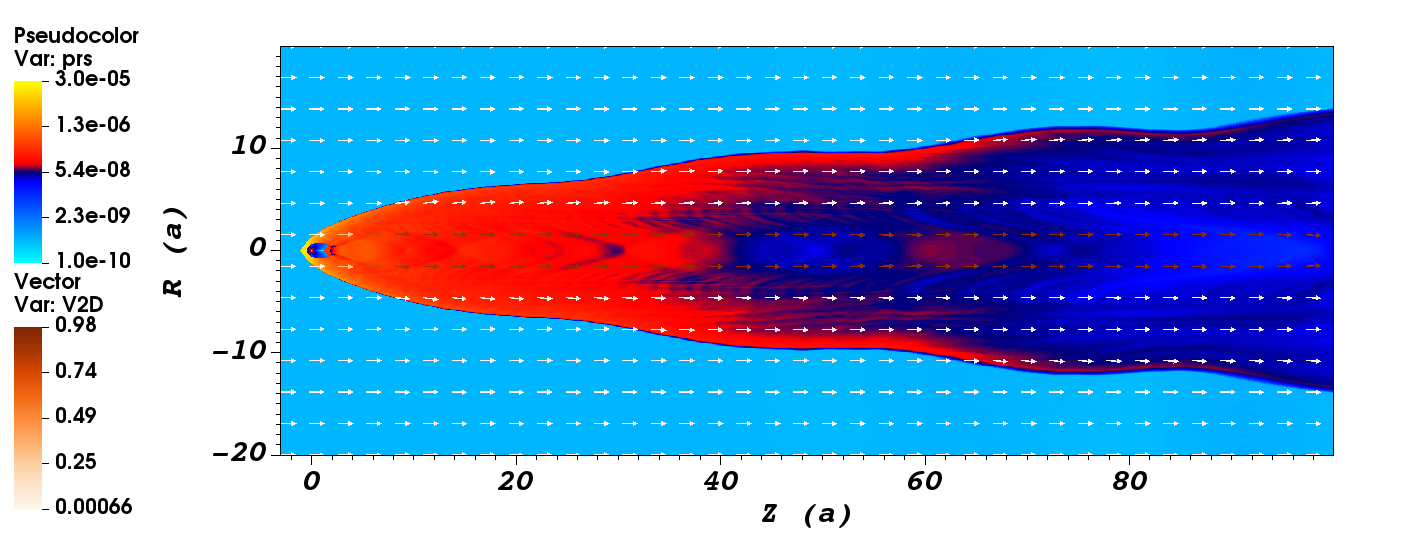}
	\caption{ Color maps of pressure (log-scale) the same models as in Fig.~\ref{fig:rho_lr}. Arrows show the velocity field, the arrow color defines the flow speed (linear scale, speed of light units).}
    \label{fig:prs_lr}
\end{figure*}

\begin{figure*}
	\includegraphics[width=\textwidth]{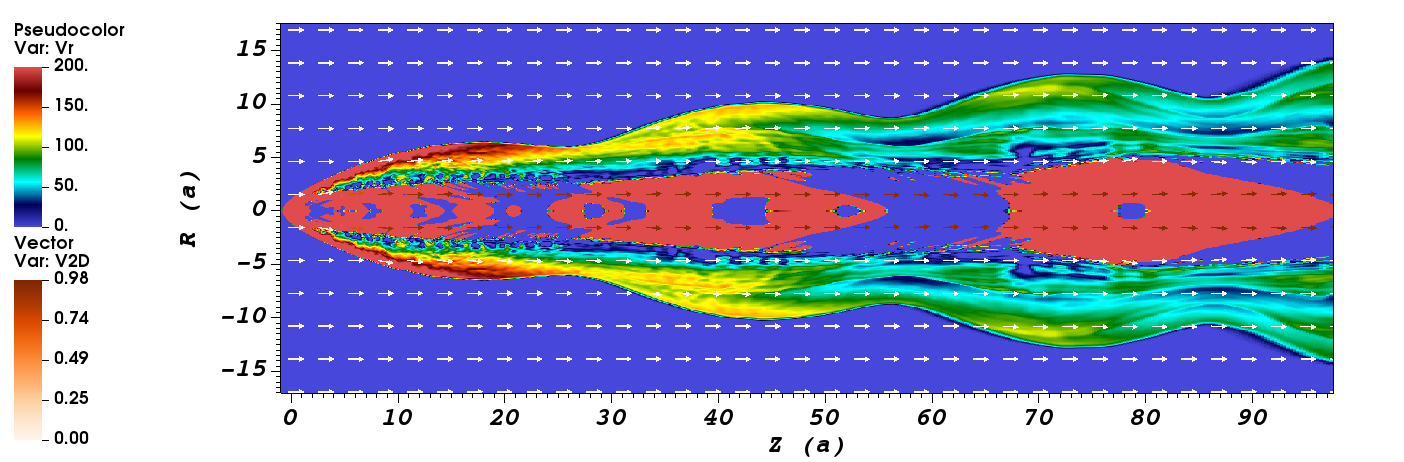}
	\includegraphics[width=\textwidth]{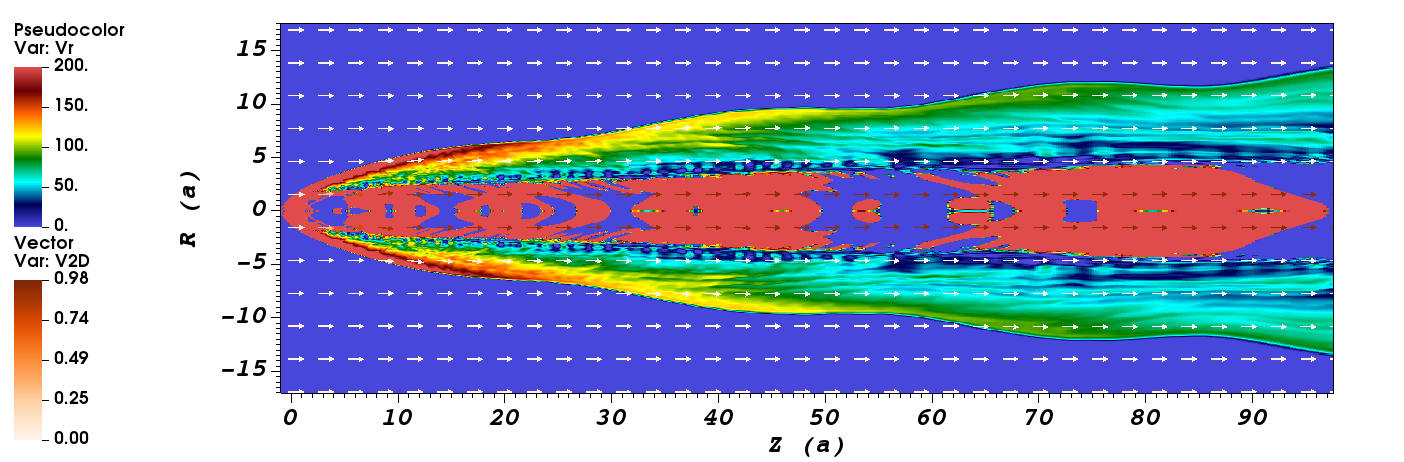}
	\caption{ Color maps of radial velocity (linear-scale, [\kms] units) for two ``low-resolution'' models with slower pulsar speed:  lr-SNEq-slow-var3 (top panel),  and lr-SNEq-slow-var2 (which has a smaller variation of \ac{ism} density, bottom panel). Arrows show the velocity field, the arrow color defines the flow speed (linear scale, speed of light units).}
    \label{fig:vr_lr}
\end{figure*}

\begin{figure*}
	\includegraphics[width=\textwidth]{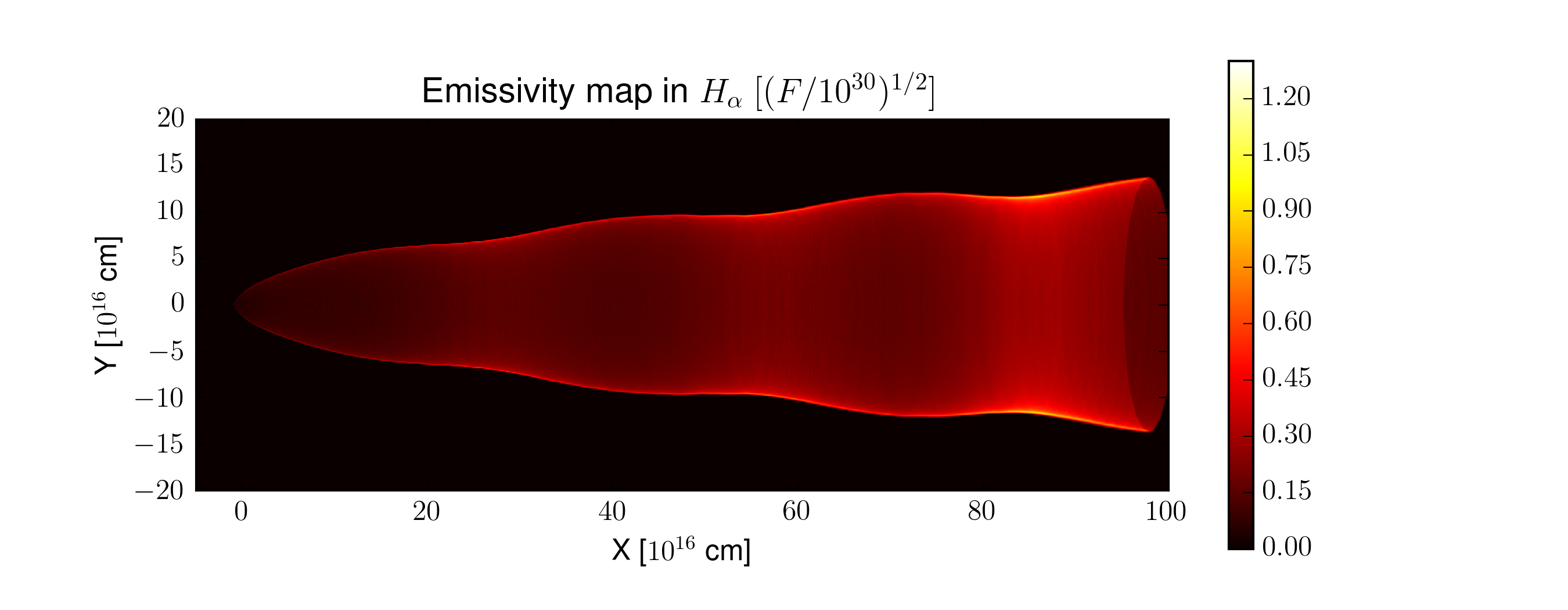}
    \caption{$H\alpha$ { line emission}  maps for model lr-SNEq-slow-var2. Viewing angle is  0.1 radian.}
    \label{fig:Ha15}
\end{figure*}

\begin{figure*}
	\includegraphics[width=\textwidth]{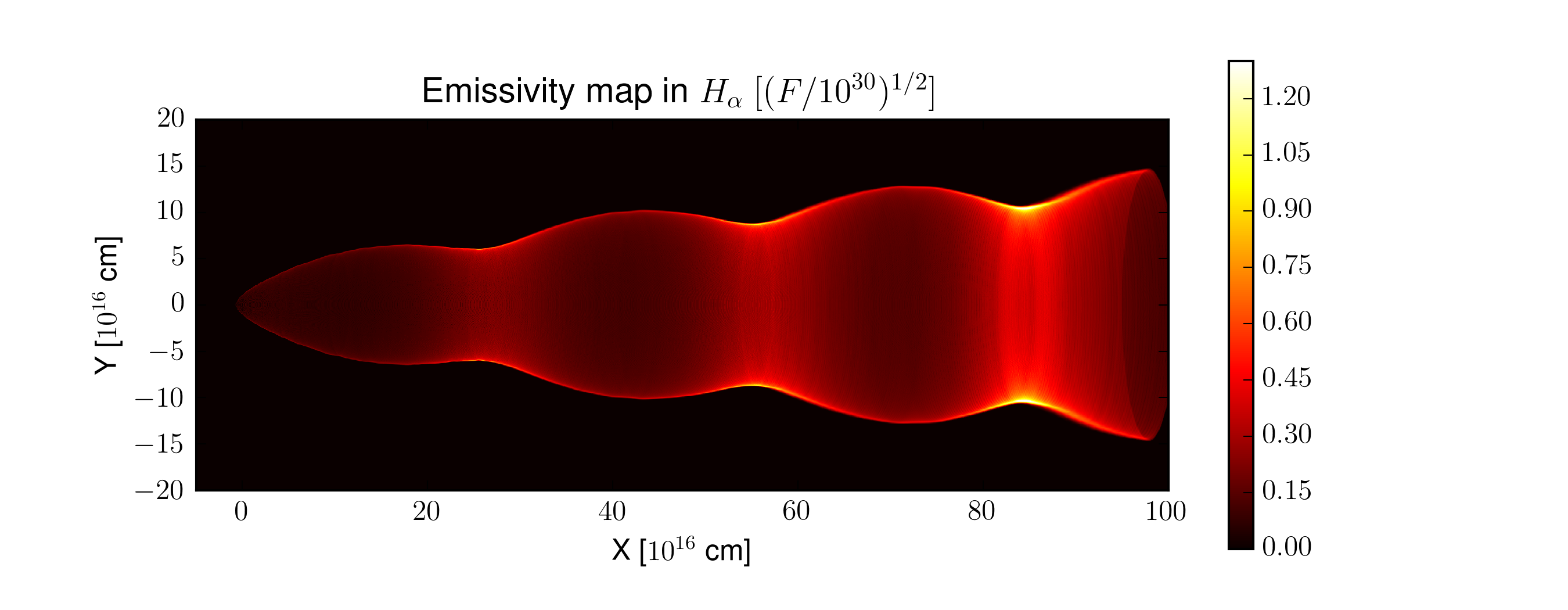}
	\includegraphics[width=\textwidth]{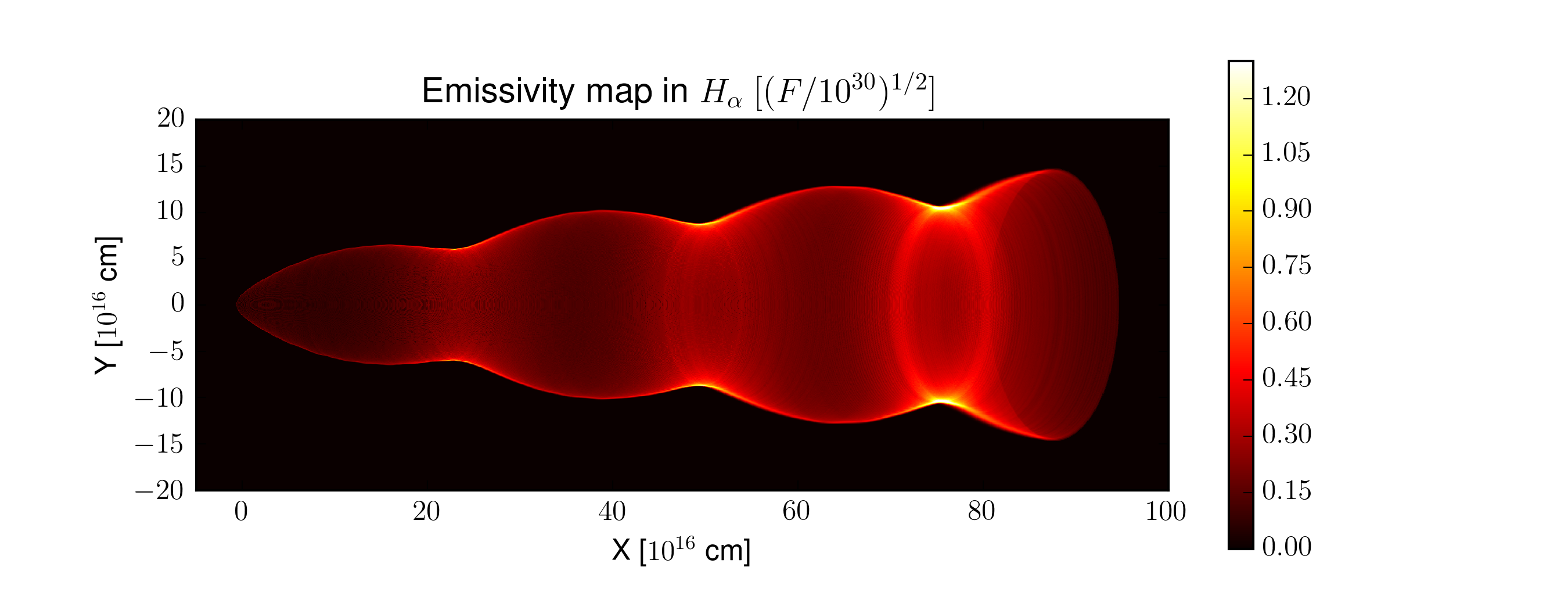}
	\includegraphics[width=\textwidth]{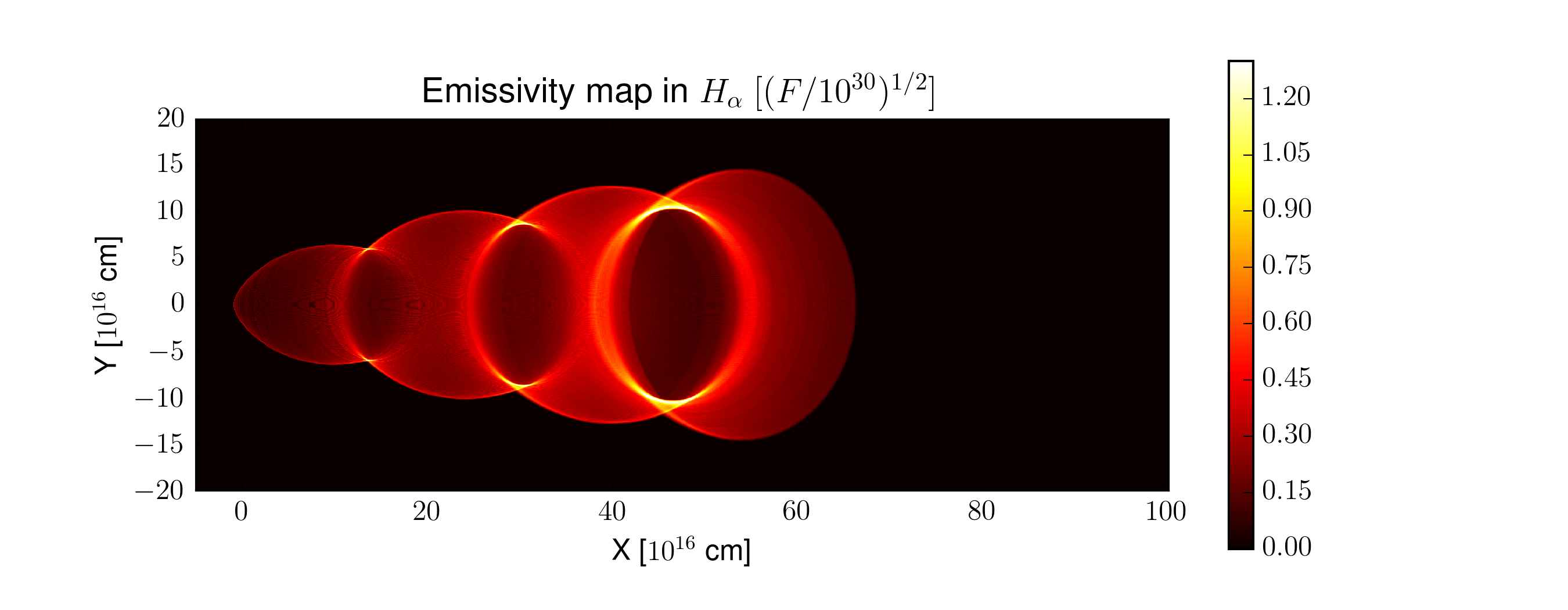}
    \caption{$H\alpha$ { line emission}  maps for the model lr-SNEq-slow-var3 and three different viewing angles: 0.1 radian (top panel), 0.5 (middle panel), and 1 (bottom panel).}
    \label{fig:Ha3}
\end{figure*}

We performed five runs with different {spacial resolutions} and model parameters to illustrate how system evolves and which parameters play a crucial role. {All the relevant parameters for each run are summarized in Table \ref{tab:models}. We name each run accordingly the following simple rule: ``run resolution (hr, lr or lr-SNEq)\footnote{"SNEq" indicates calculation of non-stationary ionization.}'' -- ``pulsar velocity (fast or slow)'' -- ``\ac{ism} spatial dependence (const or var2 or var3)''.  All the simulations were performed in the reference frame where the pulsar is at rest, and the \ac{ism} is injected from the most left border of the computational domain (which corresponds to \(z=-3a\)). At the \ac{ism} inlet the flow velocity was set to \(\varv\ns\) and the density is determined by Eq.~\eqref{eq:nISM} for \(z=-3a\). }

In the model hr-fast-const we inject  \ac{ism} matter with constant density (i.e., \(a_\rho=0\) in Eq.~\eqref{eq:nISM}) and velocity set to \(0.1c\) (see Table~\ref{tab:models}). On top panels of Fig.~\ref{fig:rho_hr} and Fig.~\ref{fig:prs_hr} distributions of density and pressure are shown. In these figures one can see the expected flow structure:  a pulsar surrounded by the unshocked pulsar wind; at larger radii a relativistic shock wave is formed; the shocked pulsar wind and \ac{ism} matter are separated by a \ac{cd} surface;  a forward shock front takes place in the \ac{ism}. On the right side from the pulsar, the shocks form a channel filled by material from the pulsar wind, which is surrounded by the shocked \ac{ism} matter.  The shock in the \ac{ism} has a very smooth conical shape. 

We see a significant growth of the \ac{kh} instability at the \ac{cd} between the shocked pulsar wind and \ac{ism}. The \ac{kh} instability triggers the formation of significant perturbations in the  shocked pulsar wind region, so \ac{ism} matter occasionally {mixes with} the shocked pulsar wind. These dense obstacles in the pulsar wind trigger the formation of additional shocks which potentially could deform the shape of the bow shock in \ac{ism}. However, the large difference (a few orders of magnitude) between the sound speed in the shocked \ac{ism} matter and the bulk speed of the shocked pulsar wind makes it almost impossible to  develop a significant deformation of the shock front in \ac{ism}. We can see the confirmation of described process {in} the upper panels of Fig.~\ref{fig:rho_hr} and Fig.~\ref{fig:prs_hr}.

The quasistationary phase of simulation for the  model hr-fast-var3 with periodic density distribution is presented {\bf in} \sout{on} the bottom panel of Fig.~\ref{fig:rho_hr} and Fig.~\ref{fig:prs_hr}. Here we injected the \ac{ism} matter  with variable density (\(a_\rho=0.5\) in Eq.~\eqref{eq:nISM}). A strong modulation of the shape of the  \ac{ism} shock can be seen. The shape of the \ac{cd} is very complicated in this simulation. We can see a series of hierarchical structures which linked to the growth of small scale vertexes triggered by \ac{kh} instability. On a larger scale, one can see elongated eddies triggered by the \ac{ism} density variation. Interesting, the impact on the crossection of the shocked pulsar wind channel is significantly smaller compared to the variation of the crossection of forward shock in \ac{ism}.  

The gas pressure, shown in Fig.~\ref{fig:prs_hr}, for the model hr-fast-const indicates on a series of more-or-less uniformly distributed shocks which are triggered in the pulsar wind channel and propagate outside in the shocked material. The exact position of the \ac{cd} is difficult to localize in this figure. The model hr-fast-var3 reveals a different behavior\sout{,}{\bf :} a series of shocks, which look like recollimation ones, can be localized just behind { the high density front} in the \ac{ism}.   

The models with slower \ac{ns} velocity require a much longer computational time, so these models were computed with a smaller computational resolution. To verify eligibility of this resolution we also performed one simulation with this resolution for a faster moving \ac{ns}. 

First, let's discuss the differences and similarities between the model lr-fast-var3 (Fig.~\ref{fig:rho_lr}) and hr-fast-var3 (Fig.~\ref{fig:rho_hr}). The shape of the forward shock in \ac{ism} is essentially the same. The main difference is the structure of the \ac{cd} surface, which in the case of lr-fast-var3 is much  smoother if compared to the one obtained in hr-fast-var3 simulations. It can be explained by the influence of the numerical viscosity which suppresses the growth of the \ac{kh} instability. { The pressure maps (Fig.~\ref{fig:prs_lr}) show similarity of the bow shock shapes, also we see similar recollimation shocks. The high pressure zones in pulsar wind follows the high density zones in the ISM.} The difference is the absence of fine structures in the shocked pulsar wind channel. In the low resolution case, the recollimation shocks have smoother structure. As we explained in \S~\ref{sec:physics} and it is shown below, the optical lines are produced at the forward shock in the \ac{ism}.  The shocked pulsar wind zone is {predominately filled with electron-positron pairs and should not produce any considerable amount of \(H\alpha\) emission}\footnote{The small scale pulsar wind zone can be seen in X-ray and radio (see \cite{2015SSRv..191..391K,2019MNRAS.484.4760B,2019MNRAS.485.2041B})}. We therefore safely claim that the resolution does not change significantly the geometry and properties of the \(H\alpha\) emitting region. 

The ``slow'' models feature a significantly larger density jump between the \ac{ism} and the pulsar wind (see Fig.~\ref{fig:rho_lr}). The \ac{cd} is more stable and the \ac{kh} instability does not disrupt the pulsar wind channel. On the other hand, the forward shock shapes are almost identical in all *-var3 models. The pressure distributions are also very similar between  lr-fast-var3 and  lr-SNEq-slow-var3 models.

The amplitude of the density variation has a strong impact on the shape of the bow shock in \ac{ism} (see Fig.~\ref{fig:prs_lr}). A variation of density by a factor of 3 (model lr-SNEq-slow-var3) creates a forward shock with a clear wavy shape, which is correlated to the \ac{ism} density profile: the higher density, the smaller { cylindrical radius of the shock}. A model with a small density variation, {lr-SNEq-slow-var2}, still features a wavy-shaped bow shock. The recollimation shocks in the pulsar wind channel are visible and have a spatial lag {relative} to the density peaks in the \ac{ism}.

In  Fig.~\ref{fig:vr_lr} we show the distribution of the {radial component of speed in cylindrical coordinates} for two models, lr-SNEq-slow-var3 and lr-SNEq-slow-var2. The radial speed at the shock decreases with the distance from the pulsar. On top of the decreasing trend, one can also see some speed increases in the low \ac{ism} density regions. These increases are mostlikely related to the change of the sound speed in \ac{ism} $c_s=\sqrt{\gamma_g p_g/\rho_g}$.  In the case of the considered models, even for ``slow'' models, the pulsar moves with record speed as compared to pulsars found in Galaxy. The ``sweet spot'' speed for $H\alpha$ line production is about $10^2 \,\kms$. {For such speed, the hydrogen ionization front is relatively thick, which makes the emission process to be efficient.   In our simulations such speeds appear only on at the edge of the computational domain (see the right part of Fig.~\ref{fig:vr_lr}). As can be seen in Figs.~\ref{fig:Ha15} and ~\ref{fig:Ha3} the $H\alpha$ line emission appears to be the most bright in this regime. In the case of a realistic \ac{ns} proper speed, the bright \(H\alpha\) region should appear close to the forward shock apex.} 

\subsection{$H\alpha$ line intensity distribution}

In Fig.~\ref{fig:Ha3} synthetic emissivity maps are shown for different viewing angles, \(\theta=0.1\), \(0.5\), and \(1\)~radian, from the top to the bottom. The high density peaks in \ac{ism} are clearly visible as bright rings. Even a relatively small variation of the \ac{ism} density leads to a very clear change of the surface brightens, with two important factors amplifying the emission: 1) total particle density; and 2) thickness of the ionization front. If the impact of the former factor is straightforward,  ($I\propto n^2$), addressing the latter one is more complicated. If the shock is too fast, the ionization take place almost instantly and {no neutral hydrogen} left to emit after shock front. If shock is too slow, hydrogen atoms do not get excited and we again do not expect any significant line emission. Therefore, to see the $H\alpha$ nebula pulsar should move with velocity $\varv\ns[7.5]\sim 1$.

The observational data \citep{2014ApJ...784..154B} show very similar features, the brighter are regions which seem to have the smaller radius, which should correspond to high \ac{ism} density  filaments crossed by the pulsar. 

If \ac{ism} density has a more complicated distribution \citep[as we see in][]{2014ApJ...784..154B}, one can expect a formation of structures which are more complex than {ring-like structures} revealed with our \ac{2d} simulations. For example, depending on the viewing angle even axial symmetric \ac{2d} distributions of emissivity can appear as bright triangles or lenses (see Fig.~\ref{fig:Ha3}, bottom). We plan to further investigate the formation of complex structures in a future study with \ac{3d} \ac{mhd} simulations.

\subsection{Guitar Nebula} 

The  parallax to pulsar B2224$+$65 measured with VLBI is $1.2^{+0.17}_{-0.2}$~mas \citep{2019ApJ...875..100D}, which corresponds to a distance of $D = 830^{+170}_{-100}$~pc. 
The observations of Guitar Nebula \citep{2002ApJ...575..407C,2016JASS...33..167D} in $H\alpha$ line revealed several depressions of the forward shock with  the angular distance of  $15''-30''$ between depression. This corresponds to a linear distance of $2-5\times10^{17}$~cm (or $\sim 0.1$~pc). A comparison of the Guitar Nebula morphology with our synthetic emissivity maps (see Fig.~\ref{fig:Ha3}) show that a modest modulation of the \ac{ism} density, $\delta\rho/\rho \approx 0.8$, can result in such shape of the forward shock. The recent observations indicate that the pulsar a few years ago passed through a region with even higher density, $\delta\rho/\rho \sim 2$ \citep{2016JASS...33..167D}. 

{The pulsar producing the Guitar Nebula moves nearly in the plane of sky. We expect that for this configuration transverse density gradients will not produce distinct morphological features. In contrast, when the line of sight is nearly aligned with the pulsar's velocity large morphological variations are expected \citep{2014ApJ...784..154B}.}

{ In the X-ray energy band, Chandra observations revealed a compact emitter spatially coinciding with pulsar B2224$+$65 and a filament structure extending by \(\sim2\)~arcmin to the north-west (i.e., making angle of \(\sim118^\circ\) with the pulsar proper velocity) \citep{2007A&A...467.1209H}. The flux corresponding to the compact emitter is \(\sim1.8\times10^{-14}\;\ergscm\) \citep{2010MNRAS.408.1216J} and its extension is limited by \(0.5''\) \citep{2007A&A...467.1209H}. For the source distance of 830 pc these parameters translates to \(L_{\rm keV}=1.5\times10^{30}\ergs\) and \(6\times10^{15}\rm\,cm\). Because of its point-like nature, the compact emitter was associated with the pulsar. Similarly to \ac{pwn} around isolated pulsar, bow-shock nebulae also feature extended non-thermal emission \citep{2017JPlPh..83e6301K}. The characteristic size of X-ray  morphology is comparable to the radius of the termination shock \citep{2019MNRAS.484.4760B}, i.e., \(\sim10^{16}\rm\,cm\) in the case of Guitar Nebula. In this region, adiabatic losses should dominate in the energy range responsible for the production of X-ray emission. According to Eq.(22) from \citet{2019MNRAS.484.4760B}, X-ray luminosity of the extended nebula should be at the level of \(0.1\%\) of the  power responsible for acceleration of the keV-emitting electrons, i.e., fainter compare to the reported pulsar emission. We therefore conclude that Chandra observatory is not expected to detect the extended emission of Guitar Nebula.}

{The extended structure in Guitar Nebula can be associated with  non-thermal particles escaping from \ac{pwn} \citep{ban08,2019MNRAS.485.2041B}. For example, \citet{ban08} argued that the X-ray feature  reacquires \ac{ism} magnetic field of $B\sim 50\rm\,\upmu G$. Large scale magnetic field can influence the shape of the termination shock. The magnetic field impact at the shock is determined by the gas normal velocity, $\varv_{s,n}$, and the Alfven speed, which is  $\varv_A\sim 10^7\rm\,cm\,s^{-1}$ (for \(B=50\rm\,\upmu G\) and plasma density of \(\rho=1/\mbox{cm}^3\)). The relative distortion of the forward shock can be estimated as $\left(\varv'-\varv_{s,n}\right)/\varv_{s,n}$, where $\varv' = \sqrt{\varv_{s,n}^2 + \varv_A^2 }$. For fast mowing pulsars, like B2224$+$65, the gas normal velocity is also very high, $\varv_{s,n}\sim \varv\ns\approx 10^8\rm\,cm\,s^{-1}$, thus the relative deformation of the forward shock remains small, \(\sim 0.01\). We therefore conclude that the suggested scenario for the formation of \(H_\alpha\) morphology in Guitar Nebula is consistent with the X-ray observations. }

\subsection{Connection to other observations}

{The Herschel satellite imaging observations of nearby molecular clouds \citep{2010A&A...518L.104M}, show that filamentary structures are characterized by length-scales with a relatively narrow distribution  around a length of $\sim0.1$~pc.  This spreads within a
factor of two for a wide range of column densities
\citep{2011A&A...529L...6A,2019A&A...621A..42A,2015MNRAS.452.3435K, 2019A&A...626A..76R}. }
{These observations
  probe  the cold component of \ac{ism}, \(T\sim 10\rm\,K\).} The properties of the hot ($T\sim10^6$~K) and warm
($T\sim10^4$~K) components of \ac{ism} are less known. {As the hot and warm components are distributed in the interstellar space between the molecular clouds, their low densities make impossible obtaining of any meaningful constrains with CO line emission, as in the case of the molecular clouds.}  {As shown above, \(H\alpha\) emission from forward shocks created by fast moving pulsars is very sensitive to the density of \ac{ism}, which allows one to obtain unique information about density structure of the warm \ac{ism}. Our interpretation of the Guitar Nebula suggests that the warm component of \ac{ism} features density filaments of the same length scale as cold \ac{ism} as revealed with Hershel observations. }

Interesting, the stripe structure seen in X-rays at the forward shock wave of \ac{snr} Tycho has a similar length scale. The characteristic angular size of $l_{gap}\sim 8''$ corresponds to linear size $\sim0.1$~pc 
\citep{2001A&A...365L.218D,2006MNRAS.369.1407K,2011ApJ...728L..28E}. { Previously, these X-ray stripes were interpreted as a result of  non-linear particle acceleration in turbulent media \citep{2011ApJ...735L..40B,2013ApJ...765L..20C}.   Based on  similarity of the linear size of X-ray features seen in Tycho and density fluctuations inferred around the Guitar Nebula, we suggest that these structures, intrinsic to the upstream medium,
are illuminated by the passage of the shock. Further down stream, strong turbulence  \citep[triggered, \eg\ by the  Relay-Taylor instability][]{2005ApJ...634..376W}  erases these structures.
}


{Similar structures might have  also been detected spectroscopically in the \ac{snr}s RX~J0852.0$–$4622 and Vela \citep{2012MNRAS.424.3145P}. Line profiles were indicating a presence of accelerating clouds, with sizes similar to the size of those inferred in the Guitar Nebula and Tycho \ac{snr}, $\sim0.1$~pc.}


We also note that the structures we inferred in the warm \ac{ism} around the Guitar Nebula should have  different origin  from the ones producing  filaments in dense
Galactic molecular clouds \citep[\eg][]{2011A&A...529L...6A}. In the case of molecular clouds the  parsec-scale filaments correspond approximately to the Jeans length.  The warm component of  ISM has much higher temperature ($\sim 10^4$~K) and respectively much larger Jeans length, $l_J\approx c_s \left( \uppi/G\rho \right)^{1/2}\sim 2 $~kpc, which significantly exceeds the revealed value. {Thus, a new mechanism operating on the same scale, $\sim 0.1$~pc, is required to explain the formation of periodic structures in the warm \ac{ism}.} 

In conclusion, we show that the dynamics and morphology of nebulae around fast moving pulsars can be used as a probe of the fine sub-parsec structure of the warm \ac{ism}. In the present paper we focused on $H\alpha$ emission, however our approach can be generalized to other prominent optical lines. {Numerical simulation of synthetic emission maps for several ``optical'' lines (from UV  to IR bands)  can provide unique information about the properties of \ac{ism}.   Future development of such methods also can deliver independent measurements of proper pulsar velocities, \ac{ism} properties and its chemical composition. }

\section*{Acknowledgements}

We appreciate R.J. Tuffs and Dmitri Wiebe for useful discussion. 
The calculations were carried out in the CFCA cluster of National Astronomical Observatory of Japan.
We thank the {\it PLUTO} team for the opportunity to use the {\it PLUTO} code. 
The visualization of the results of \ac{mhd} simulations were performed in the VisIt package \citep{HPV:VisIt}. We also acknowledge usage of {\it PyCUDA} and {\it matplotlib} libraries.  
ML would like to acknowledge support by  NASA grant 80NSSC17K0757 and NSF grants 10001562 and 10001521.
DK is supported by JSPS KAKENHI Grant Numbers JP18H03722, JP24105007, and JP16H02170.



\bibliographystyle{mnras}
\bibliography{ismpw}



\appendix

\section{GPU simulation of the \(H\alpha\) synthetic maps}
\label{sec:Halpha}
To obtain \(H\alpha\) brightness maps the emission coefficient, which was obtained from \ac{mhd} simulations, is to be
integrated over line of sight at the postprocessing stage. We were interested in a very efficient algorithm, as for each
\ac{mhd} simulation we studied the influence of the direction of the pulsar velocity and time evolution of the synthetic
maps, which is important for understanding the formation of the visual structures seen in \(H\alpha\) emission. We
therefore adopted an approach based on the GPU computing. We used {\it PyCUDA}, a {\it CUDA} API implementation for {\it
  python}. The {\it python} script consists of several stages, which (i) read the data from \ac{mhd} simulations; (ii)
process a template for {\it CUDA C} code; (iii) copy the data to the GPU device and run block of GPU processes, and (iv)
retrieve synthetic map from the device and create a visual file using {\it matplotlib}. A version of the script that
includes stages (ii)-(iv), \ie computes a synthetic map for a test array of emissivity is available at ``https://github.com/dmikha/GPUmaps.git''. Below we
briefly outline the key steps in this fairly simple script.

We first start with general explanation of the used algorithms. The \ac{mhd} box coordinates do not match the ``real
world'' Cartesian coordinates (where the synthetic maps are computed: \(XY\) is the plane of sky and \(Z\) axis directed
along line-of-sight), and the ``real world'' coordinates need to be transforms to the \ac{mhd} box coordinates. As the
obtained coordinates do not match the nodes of the \ac{mhd} grid, one needs to assign some emission coefficient at that
specific location. This requires obtaining nearby nodes from the \ac{mhd} array and approximating the value. To obtain
the nearby nodes we used a recursive high-efficiency algorithm suitable for monotonic arrays. To determine the
emission coefficient at the revealed location, we used a bilinear approximation method. This resulted in a sectionally
linear dependence of the emission coefficient, thus any high-order integration methods are not justified for the
integration over the line-of-sight. We therefore utilized the trapezoidal integration rule. 

There are detailed tutorial how to use {\it PyCUDA} here we just provide short comments on our script. The script should
include listing of the {\it CUDA C} code for the main computational block. This code is process by {\it nvcc} compiler, \eg
\begin{verbatim}
mod = pycuda.compiler.SourceModule(nvcc_code)
\end{verbatim}
Before the compilation, however, one can modify the {\it nvcc} code with various {\it python} tools, \eg, to define values of some parameters:
\begin{itemize}
\item {\it nvcc} code contains
\begin{verbatim}
...
#define s_R $s_R
...
\end{verbatim}
\item {\it python} script contains
\begin{verbatim}
...
nvcc_code = nvcc_listing.substitute(s_R  = N_r)
...
\end{verbatim}
\end{itemize}

An instance of {\it nvcc} function callable from python can be created \eg with
\begin{verbatim}
create_map_cuda = mod.get_function("create_map")
\end{verbatim}
After that one needs to transfer data to the device memory. In {\it PyCUDA} there are various ways of doing this, in our script we used two slightly different syntaxes for that. In the first case we transfer the auxiliary array with
\begin{verbatim}
...
mod = SourceModule(nvcc_code)
R_d = mod.get_global('R_d')[0]
pycuda.driver.memcpy_htod(R_d, r.astype(np.float32))
...
\end{verbatim}
In the second case we used the data transfer at the function call:
\begin{verbatim}
...
map = np.zeros((Ny_grid,Nx_grid)).astype(np.float32)
create_map_cuda(
    pycuda.driver.Out(map), 
    block=my_block, grid=my_grid) 
...
\end{verbatim}
Here ``my\_block'' and ``my\_grid'' are tuples for the processes run on GPU, and should be selected according the capacity of the device. The remaining parts of the code are essentially usual {\it python} or {\it C} codes.


\bsp	
\label{lastpage}
\end{document}